\documentclass[aps,prb,10pt,twocolumn,eqsecnum]{revtex4}
\usepackage{graphicx,amsmath,amssymb,bm,bbm,latexsym}

\usepackage{color}
\allowdisplaybreaks

\newcommand{\be}{ \begin{equation}}
\newcommand{\ee}{ \end{equation}}

\usepackage{color}

\newcommand{\SU}[1]{\textcolor{black}{#1}}

\usepackage{ulem}

\begin{document}

\title{
Effects of Zeroline and Ferrimagnetic Fluctuation \\ on Nuclear Magnetic Resonance for Dirac Electrons 
\\ in Molecular Conductor $\alpha$-(BEDT-TTF)$_2$I$_3$} 

\author{A. Kobayashi, Y. Suzumura
}

\affiliation{
Department of Physics, Nagoya University, Furo-cho, Chikusa-ku, Nagoya, 464-8602 Japan
}

\date{\today}

\begin{abstract}
We re-examine the wave function of two-dimensional massless Dirac electron in $\alpha$-(BEDT-TTF)$_2$I$_3$ consisting of four molecules A, A', B and C in a unit cell,  using a tight-binding model.
We find zerolines \SU{in the Brillouin zone,  on which
the component  of the wave function becomes  zero 
for $B$ or $C$ sites. 
The zerolines, which are  bounded by two Dirac points at $\pm {\bf k}_0$ 
and pass through the $M$- or $Y$-points, 
result in a fact that} the density of states of 
the B site exhibits 
 no the Van Hove singularity  near the energy of the Dirac points. 
\SU{By taking account of the on-site Coulomb interaction
within the random phase approximation,}
we examine the spin fluctuation 
in order to investigate properties of the nuclear magnetic resonance for temperatures $T > 50$K.
In the region for $100 < T < 300$K, 
it is shown that the Knight sift for $B$-site monotonously decreases  with decreasing temperature, 
owing to lack of the Van Hove singularity, 
while it shows a maximum  
for the other sites ($A$, $A^\prime$ and $C$ sites).
In the region for $50 < T < 100$K, it is shown that 
the Knight sift is convex downward and the Korringa ratio increases with decreasing temperature for $B$-site.
Such a behavior originates from the ferrimagnetic spin fluctuation related to the zerolines.
These results are consistent with those of the nuclear magnetic resonance experiments.
\end{abstract}


\maketitle

\section{Introduction}

Molecular conductor $\alpha$-(BEDT-TTF)$_2$I$_3$
\cite{TajimaRev2009} 
has highly two-dimensional electronic system in the plane of BEDT-TTF$^{1/2}$ molecule 
owing to layered structure with the plane of I$_3^-$ anion, 
and has brought
 much interest by the variety of electronic states, such as a charge ordered state
\cite{SeoRev2004}, a superconducting state in the presence of charge ordering, 
and a zero gap state (ZGS) with a massless Dirac electron \cite{KobayashiRev2009}.

The charge ordered state was suggested theoretically 
 using an extended Hubbard model \cite{Kino-Fukuyama,Seo,Hotta},  
and was confirmed by NMR experiment \cite{TakahashiStripe}.
The superconducting state \SU{with the charge ordering} 
under uniaxial pressure along the stacking axis ($a$-axis)\cite{Tajima-Ebina2002}was also investigated theoretically using the extended Hubbard model
\cite{Kobayashi2005}.
A narrow gap state (NGS) was suggested to explain both anomalous increase of Hall coefficient and almost-constant resistivity with decreasing temperature 
at high pressures \cite{KajitaFirst}.
\SU{From the calculation of the NGS, it was obtained that 
 the charge gap disappears at high pressure, leading to the density of state (DOS) vanishing  linearly  at the Fermi energy
\cite{Kobayashi2004}.
Furher, the ZGS with a Dirac cone in energy dispersion} was found theoretically \cite{Katayama2006ZGS} 
using the transfer energies of ref. \onlinecite{Kondo2005}, 
and was also confirmed by the first principle calculations \cite{Kino,Ishibashi}.
 The energy spectrum near the Fermi energy \SU{exhibits  two tilted Dirac cones, which} are described by the tilted Weyl equation \cite{Kobayashi2007,Montambaux2008TiltedWeyl}.
 The tilt of the Dirac cone has been 
confirmed by a comparison between the theoretical and experimental
results for the temperature ($T$) dependence of
the Hall coefficient \cite{Kobayashi2008,Tajima2009}
and the angular dependence of the magnetoresistance \cite{Tajima2009,Morinari2009}.
\SU{ From the calculation of the variation of the Dirac point, the emergence of a pair of massive Dirac points is predicted  in the charge ordering state  at low pressures\cite{KobayashiPRB2010},  the merging of two massless Dirac electrons is shown at extremely high pressures \cite{Kobayashi2007}, 
 suggesting} robustness of Dirac electrons against uniaxial pressures.
\SU{Effects of electron correlation for long range Coulomb repulsion} 
  in the tilted Dirac electron system have been investigated 
theoretically in the absence of magnetic field \cite{Nishine2010,Nishine2011} and 
in the presence of magnetic field \cite{Kobayashi2009QHE} 
.
\SU{As for the short range part of Coulomb repulsion, on the other hand, 
effect of fluctuations in ZGS has not yet investigated theoretically} 
in the context of inequivalence of BEDT-TTF sites 
(${\rm A} ={\rm A}^\prime \ne {\rm B} \ne {\rm C}$ as shown in Fig. 1).

\SU{The property of the wave function plays important roles in the DOS and 
the electron-hole excitation in $\alpha$-(BEDT-TTF)$_2$I$_3$.
The anomalous behavior of the Bloch wave functions exists 
in the vicinity of the Dirac point. 
 The momentum dependences of the velocity matrix and the charge density 
exhibit a singularity at the point.
\cite{Kobayashi2007}
The angular dependence of the wave functions for each site 
  reveals a fact that the absolute value   becomes zero for  
${\rm B}$ site or  ${\rm C}$ site  at a special direction
 of the point
\cite{Katayama2009}.
The theoretical calculation of Knight shift $K_\alpha$ and $(1/T_1 T)_\alpha$
\cite{Katayama2009}.
 has been performed based on a tight-binding model for $\alpha$-(BEDT-TTF)$_2$I$_3$ where 
 $K_\alpha \propto T$ and $(1/T_1 T)_\alpha \propto T^2$ 
owing to the Dirac cone spectrum.
The Knight shift for the respective site reveals a relation, $K_{\rm C} > K_{\rm A} > K_{\rm B}$, which is consistent with experimental results,
\cite{TakahashiNMR} 
 originates from both the tilt of Dirac cone and property of wave function in the vicinity of the Dirac point.
Thus, the inequivalence of BEDT-TTF sites observed in NMR experiments 
  reveals inner degree of freedom of Dirac electron in molecular conductors.
\cite{TakahashiNMR,KanodaNMR1,KanodaNMR2} 
}

\SU{However, the tight binding model is not enough to explain the details, e.g, the non-linear temperature dependences of $K_\alpha$ as shown in the following experiment.} 
In high temperature region ($T > 100$K), 
the Knight shift $K_\alpha$ with decreasing temperature
monotonously decreases for $\alpha ={\rm B}$ site, 
while it exhibits a maximum for $\alpha ={\rm A}$ and ${\rm C}$ sites.
In medium temperature region ($50 < T < 100$K), 
 the Knight shift is convex downward with decreasing temperature for ${\rm B}$ site, 
while the components for ${\rm A}$ and ${\rm C}$ sites exhibit linear $T$-dependences.
\cite{TakahashiNMR,KanodaNMR2} 
\SU{For a local NMR relaxation rate $1/T_1$,
 the difference in  $(1/T_1 T)_\alpha$ in  $\alpha ={\rm A}$, ${\rm B}$, and ${\rm C}$ sites,  is small 
 in low temperature region ($T < 50$K), 
 i.e.,} all components are convex downward with 
 $T^2$-dependences for $T < 100K$. 
As a result, the Korringa ratio $(1/T_1 T K^2)_\alpha$  
for $\alpha ={\rm B}$ site clearly increases with decreasing $T$ 
in medium temperature region, and exhibits an inequality 
$(1/T_1 T K^2)_{\rm B} > (1/T_1 T K^2)_{\rm A} > (1/T_1 T K^2)_{\rm C}$
\cite{KanodaNMR2} .

In the present paper, we re-examine the wave function in $\alpha$-(BEDT-TTF)$_2$I$_3$ 
using a tight-binding model where the transfer integrals are given by 
 the first principle calculation \cite{Kino}, 
and calculate the spin fluctuation within the random phase approximation 
on the on-site Coulomb interaction to investigate 
$K_\alpha$ and $(1/T_1 T)_\alpha$ for $T > 50$K.
\SU{Based on the formulation in \S II, we demonstrate following new results 
in \S III.} 
We find zerolines in the two-dimensional Brillouin zone, where the wave function is zero for the components of $B$ or $C$ site.  They are bounded by the two Dirac points at $\pm {\bf k}_0$. 
The zerolines passing through the $M$- or $Y$-points 
 give rise to the absence of  two Van Hove singularities \SU{for only the $B$-site.}
In the high temperature region for $100 < T < 300$K, 
the present numerical results on $T$-dependences of $K_\alpha$, 
are consistent with the experimental results \cite{TakahashiNMR,KanodaNMR1,KanodaNMR2}, 
where existence (absence) of Van Hove singularities for $\alpha ={\rm A}$ and ${\rm C}$ sites 
($\alpha ={\rm B}$ site) is essential.
In the region for $50 < T < 100$K, it is shown that 
$K_{\rm B}$ is convex downward, $(1/T_1 T K^2)_{\rm B}$ increases with decreasing $T$, 
and then the inequality 
$(1/T_1 T K^2)_{\rm B} > (1/T_1 T K^2)_{\rm A} > (1/T_1 T K^2)_{\rm C}$ is reproduced, 
owing to the ferrimagnetic spin fluctuation related to the zerolines and enhanced by 
the on-site Coulomb interaction.
These results are also consistent with those experimental results
\cite{TakahashiNMR,KanodaNMR1,KanodaNMR2}.
In \S IV, summary and discussion are given.

\section{FORMULATION }

\begin{figure}
\includegraphics[height=60mm]{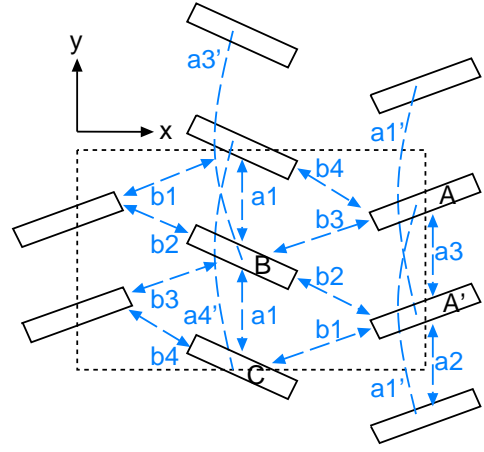}
\caption{\label{unitcell}
The model describing the electronic system of $\alpha$-(BEDT-TTF)$_2$I$_3$ \cite{Kino,Mori1984,Mori1999}.
The unit cell consists of four BEDT-TTF molecules ${\rm A}$, ${\rm A}^\prime$, ${\rm B}$ and ${\rm C}$ with  ten transfer energies.
The $a$- and $b$-axis in the conventional notation correspond to the $y$- and $x$-axis in the present paper.
The molecules ${\rm A}$ and ${\rm A}^\prime$ are equivalent in the presence of the inversion symmetry 
in the ZGS. (Color Online)
}
\end{figure}

The model describing the two-dimensional electronic system in
$\alpha$-(BEDT-TTF)$_2$I$_3$ is shown in
Fig.~\ref{unitcell} \cite{Kino,Mori1984,Mori1999}. 
The unit cell consists of four BEDT-TTF molecules
   on sites A, ${\rm A}^\prime$, B and C, 
where A is equivalent to ${\rm A}^\prime$ so that inversion symmetry is preserved, 
while the sites A, B and C are inequivalent.
There are six electrons for the four molecules in a unit cell, 
thus the bands are $3/4$-filled. 
On the basis of the HOMO orbitals of these sites \cite{Kino-Fukuyama,Seo}, 
these electrons are described by a Hubbard model with the on-site Coulomb interaction $U$,  
\begin{eqnarray}
H &=& \sum_{( i \alpha : j \beta ),
\sigma}
 (t_{i \alpha; j \beta}\
   a^{\dag}_{i\alpha\sigma}a_{j\beta\sigma}+ {\rm h. c.} )
     \nonumber\\
 &+& \sum_{i\alpha}
    U\ a^{\dag}_{i\alpha\uparrow}a^{\dag}_{i\alpha\downarrow}
      a_{i\alpha\downarrow}a_{i\alpha\uparrow} ,
\label{h1}
\end{eqnarray}
where $i, j$ denote indices of a given unit cell,
   and $\alpha, \beta (={\rm A}$, ${\rm A}^\prime$, ${\rm B}$ and ${\rm C}$) are 
 indices of BEDT-TTF sites in the unit cell.
In the first term,
   $a^{\dag}_{i\alpha\sigma}$ denotes a creation operator
    with spin $\sigma (=\uparrow ,\downarrow)$ and
    $t_{i\alpha ; j \beta}$ is the transfer
       energy between the $(i,\alpha)$ site and the  $(j,\beta)$ site.
Throughout the paper, $\hbar$ and the lattice constant $a$ are taken as unity.
Hereafter, the energies are given in eV, and 
the temperature is also given by $k_{\rm B} T$ in eV, 
{\it i. e.} $1$eV =$10^4$K, where $k_{\rm B}$ is the Boltzmann factor. 

The transfer energies at finite temperature $T$ is estimated by the interpolation formula
\cite{Katayama2009}
\begin{equation}
t_X (T) = t_X ({\rm LT}) +(t_X ({\rm RT}) - t_X ({\rm LT}))(T-{\rm LT})/({\rm RT} - {\rm LT})  .
\end{equation}
The transfer energies $t_X ({\rm RT})$ and $t_X ({\rm LT})$ 
are given by the 1st principle calculation \cite{Kino} , where 
$
t_{a1}({\rm RT}) = -0.0101, 
t_{a2}({\rm RT}) = -0.0476,
t_{a3}({\rm RT}) = 0.0093, 
t_{b1}({\rm RT}) = 0.1081,
t_{b2}({\rm RT}) = 0.1109, 
t_{b3}({\rm RT}) = 0.0551, 
t_{b4}({\rm RT}) = 0.0151,
t_{a1^\prime}({\rm RT}) = 0.0088, 
t_{a3^\prime}({\rm RT}) = 0.0019,
t_{a4^\prime}({\rm RT}) = 0.0009, 
$ and 
$
t_{a1}({\rm LT}) = -0.0267, 
t_{a2}({\rm LT}) = -0.0511,
t_{a3}({\rm LT}) = 0.0323, 
t_{b1}({\rm LT}) = 0.1241,
t_{b2}({\rm LT}) = 0.1296, 
t_{b3}({\rm LT}) = 0.0513, 
t_{b4}({\rm LT}) = 0.0512,
t_{a1^\prime}({\rm LT}) = 0.0119, 
t_{a3^\prime}({\rm LT}) = 0.0046,
t_{a4^\prime}({\rm LT}) = 0.0060, 
$ , where low temperature ${\rm LT}=0.0008$ and we put room temperature ${\rm RT}=0.03$.

The Hamiltonian is diagonalized numerically for a given ${\bf k}$ in each spin subspace, 
according to

\begin{eqnarray}
\sum_{\beta=1}^{4}
\tilde{\epsilon}_{\alpha\beta\sigma}({\bf k})\ d_{\beta \gamma \sigma}({\bf k})
&=&\xi_{\gamma \sigma}({\bf k})\ d_{\alpha \gamma \sigma}({\bf k}) ,
\label{eigenvalue}
\end{eqnarray}
where $\xi_{\gamma \sigma}$ are the eigenenergies ordered such that,
$
 \xi_{1 \sigma}({\bf k}) > \xi_{2 \sigma}({\bf k})
  > \xi_{3 \sigma}({\bf k}) > \xi_{4 \sigma}({\bf k})
$  ($\gamma=1,2,3,4$  is the  band index), and $d_{\alpha \gamma \sigma}({\bf k})$  are
the corresponding eigenvectors.
The $T$-dependence of the chemical potential owing to the electron-hole asymmetry 
\cite{Kobayashi2008} is taken into account in the present calculation, 
although it has not been considered in \onlinecite{Katayama2009}.

The bare susceptibility on the site representation is given by\cite{Kobayashi2004} 
\begin{eqnarray}
&&[ \hat{\chi}^0 ]_{\alpha \beta} = \chi_{\alpha \beta}^0 ({\bf q}, \omega_l ) \nonumber \\
&=& -\frac{T}{N_L} \sum_{{\bf k}n} 
G_{\alpha \beta} ({\bf k}+{\bf q}, \epsilon_n +\omega_l )
G_{\beta \alpha} ({\bf k}, \epsilon_n ) \\
&=& -\frac{T}{N_L} \sum_{{\bf k}n} 
\frac{F_{\alpha \beta} ({\bf k} ,{\bf q})}
{({\rm i}(\epsilon_n +\omega_l ) - \xi_\gamma ({\bf k} +{\bf q} ))
({\rm i}\epsilon_n - \xi_\gamma ({\bf k}))} ,
\end{eqnarray}
where the bare Green function on the site representation is defined by 
\begin{equation}
G_{\alpha \beta} ({\bf k}, \epsilon_n ) \equiv \sum_\gamma 
d_{\alpha \gamma} ({\bf k})  d_{\beta \gamma}^\ast ({\bf k}) 
/({\rm i}\epsilon_n - \xi_\gamma ({\bf k}))
\end{equation}
with $\epsilon_n =(2n+1) \pi T$ and $\omega_l = 2n \pi T$ are the Matsubara frequencies, 
and the form factor is defined by
\begin{eqnarray}
F_{\alpha \beta} ({\bf k} ,{\bf q}) &=& F_{\alpha \beta}^{(1)} ({\bf k} ,{\bf q}) 
+ F_{\alpha \beta}^{(2)} ({\bf k} ,{\bf q}) , \\
F_{\alpha \beta}^{(1)} ({\bf k} ,{\bf q}) &=& \sum_{\gamma =1,2} F_{\alpha \beta}^{\gamma \gamma} ({\bf k} ,{\bf q}) , \\
F_{\alpha \beta}^{(2)} ({\bf k} ,{\bf q}) &=& \sum_{\gamma =1,2} F_{\alpha \beta}^{\gamma \bar{\gamma}} ({\bf k} ,{\bf q}) ,  \\
F_{\alpha \beta}^{\gamma \gamma^\prime} ({\bf k} ,{\bf q}) &=& 
 d_{\alpha \gamma} ({\bf k}+{\bf q}) d_{\beta \gamma} ({\bf k}+{\bf q})^\ast
 d_{\beta \gamma^\prime} ({\bf k}) d_{\alpha \gamma^\prime} ({\bf k})^\ast , 
\nonumber \\
\end{eqnarray}
with $\bar{\gamma} \ne \gamma$ indicating another band within the conduction ($\gamma =1$) and valence ($\gamma =2$) bands.
The form factor reflects the character of each site for 
the intraband ($F_{\alpha \beta}^{(1)} ({\bf k} ,{\bf q})$) 
and \SU{interband} ($F_{\alpha \beta}^{(2)} ({\bf k} ,{\bf q}) $) fluctuations. 

In the presence of spin symmetry, the longitudinal spin susceptibility, $\hat{\chi}^s$, 
and the transverse spin  susceptibility, $\hat{\chi}^\pm$, are given by\cite{Kobayashi2004}  
\begin{equation}
\hat{\chi}^s = \hat{\chi}^\pm = (1-\hat{\chi}^0 \hat{U})^{-1} \hat{\chi}^0 
\end{equation}
with $\hat{U} =U \hat{I}$ and $\hat{I}$ is the unit matrix.

The knight shift $K_\alpha$ (the local spin susceptibility) 
and the local NMR relaxation rate $(1/T_1)_\alpha$ 
with ${\rm A}$, ${\rm A}^\prime$, ${\rm B}$ and ${\rm C}$ sites 
are given by \cite{Katayama2009}
\begin{eqnarray}
K_\alpha &=& \sum_\beta \chi_{\alpha \beta}^s ({\bf 0}, 0 ) ,\\
(1/T_1)_\alpha &=& T \lim_{\omega \rightarrow 0} \sum_{\bf q} 
{\rm Im} [\chi_{\alpha \alpha}^\pm ({\bf q}, \omega )]/\omega . 
\end{eqnarray}

\section{Results}

\subsection{Zeroline and density of states}

Figure 2(a), (b), (c) and (d) show absolute values of the wave functions 
of the conduction band ($\gamma =1$) 
for the components of $B$ site (Figs 2(a) and (b)) and $C$ site (Figs 2(c) and (d)).
We find "zerolines" which are curved lines where absolute value of the wave function is zero, 
only for the components of $B$ and $C$ sites.
The zerolines for both $B$ and $C$ sites are bounded by two Dirac points at $\pm {\bf k}_0$ in the Brillouin zone, 
 \SU{where the wave functions are discontinuous}.
In the conduction band, 
the zeroline for $B$ ($C$) site passes through the $M$- ($Y$-)point as shown in Fig 2(e).
In the valence band, on the contrary, 
the zeroline for $B$ ($C$) site passes through the $Y$- ($M$-)point.
The zeroline for $B$ ($C$) site in conduction band coincides with 
the zeroline for $C$ ($B$) site in valence band in the vicinity of the Dirac points.
, while these are slightly different far from the Dirac points, 
since the conduction and valence bands are not equivalent in $\alpha$-type organic conductors.

\begin{figure}
\includegraphics[height=30mm]{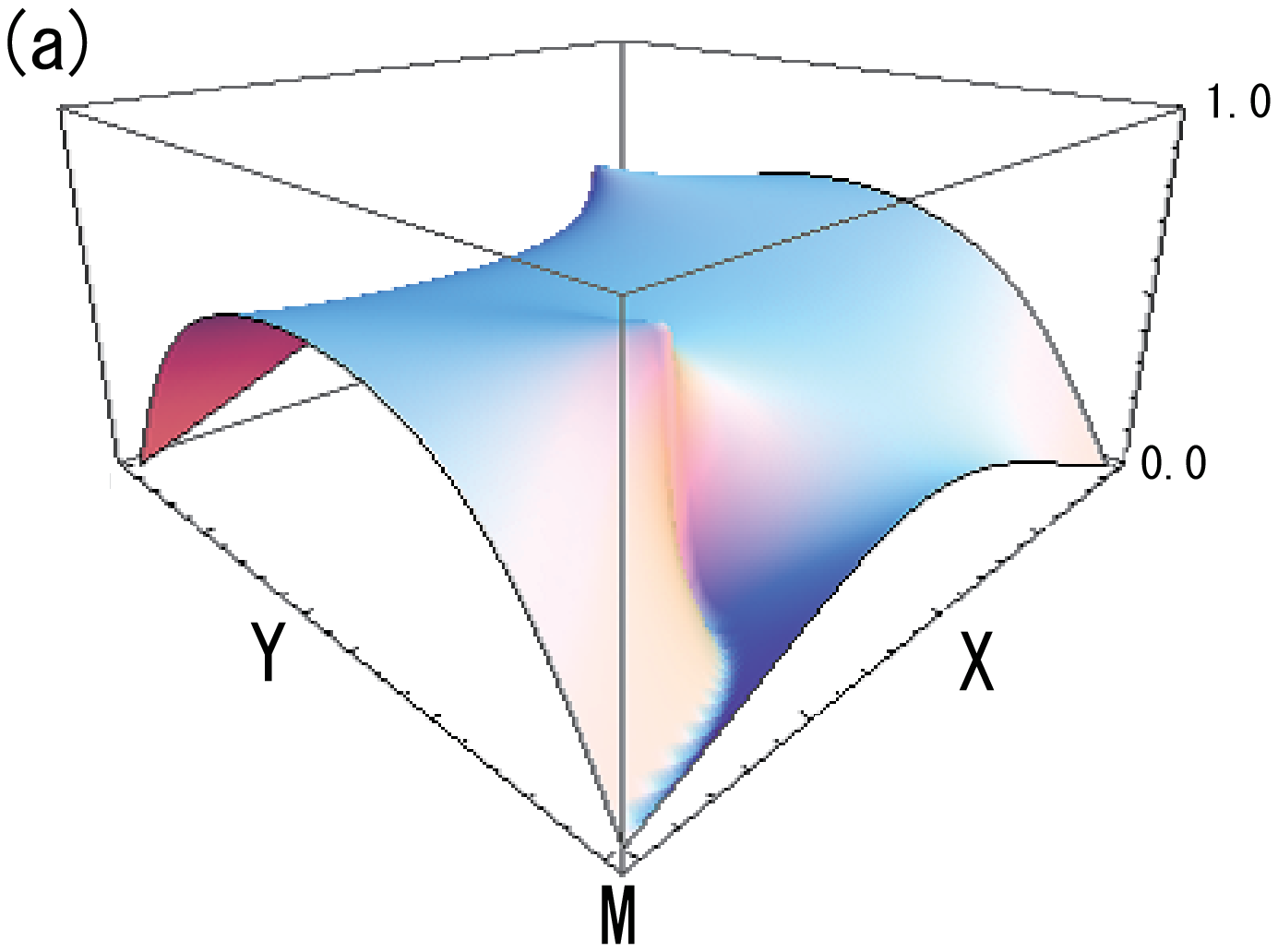}
\includegraphics[height=40mm]{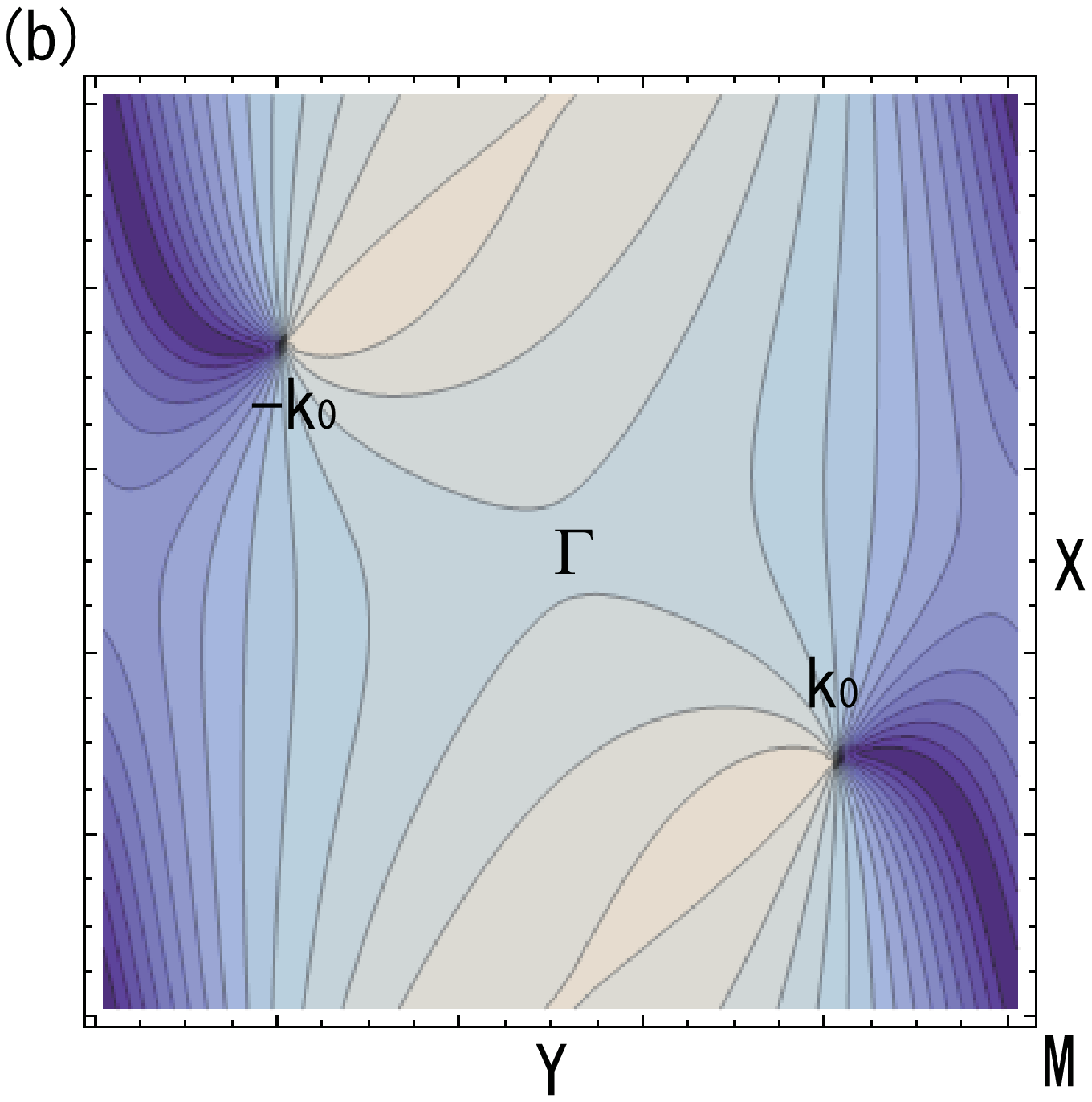}
\includegraphics[height=30mm]{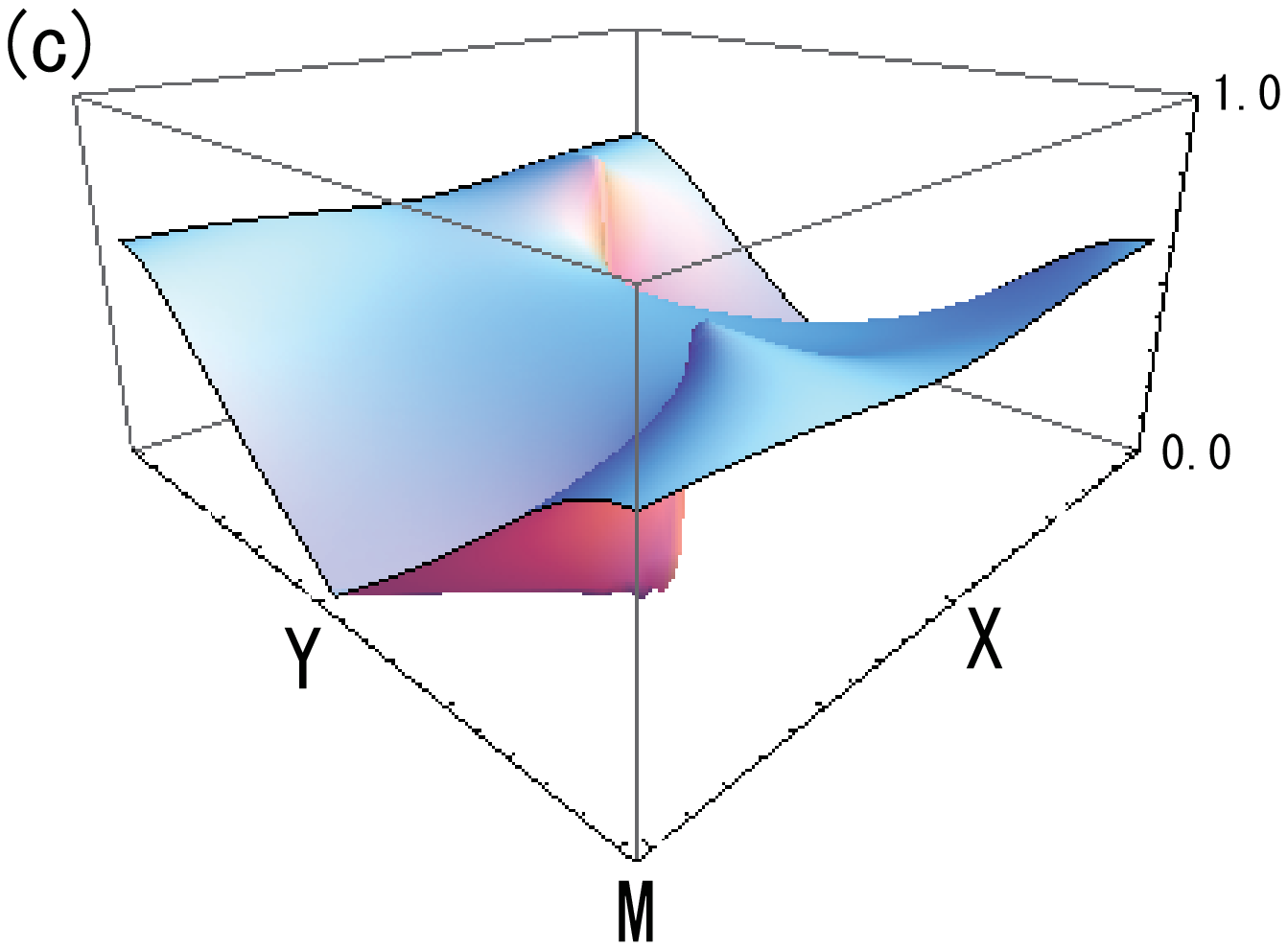}
\includegraphics[height=40mm]{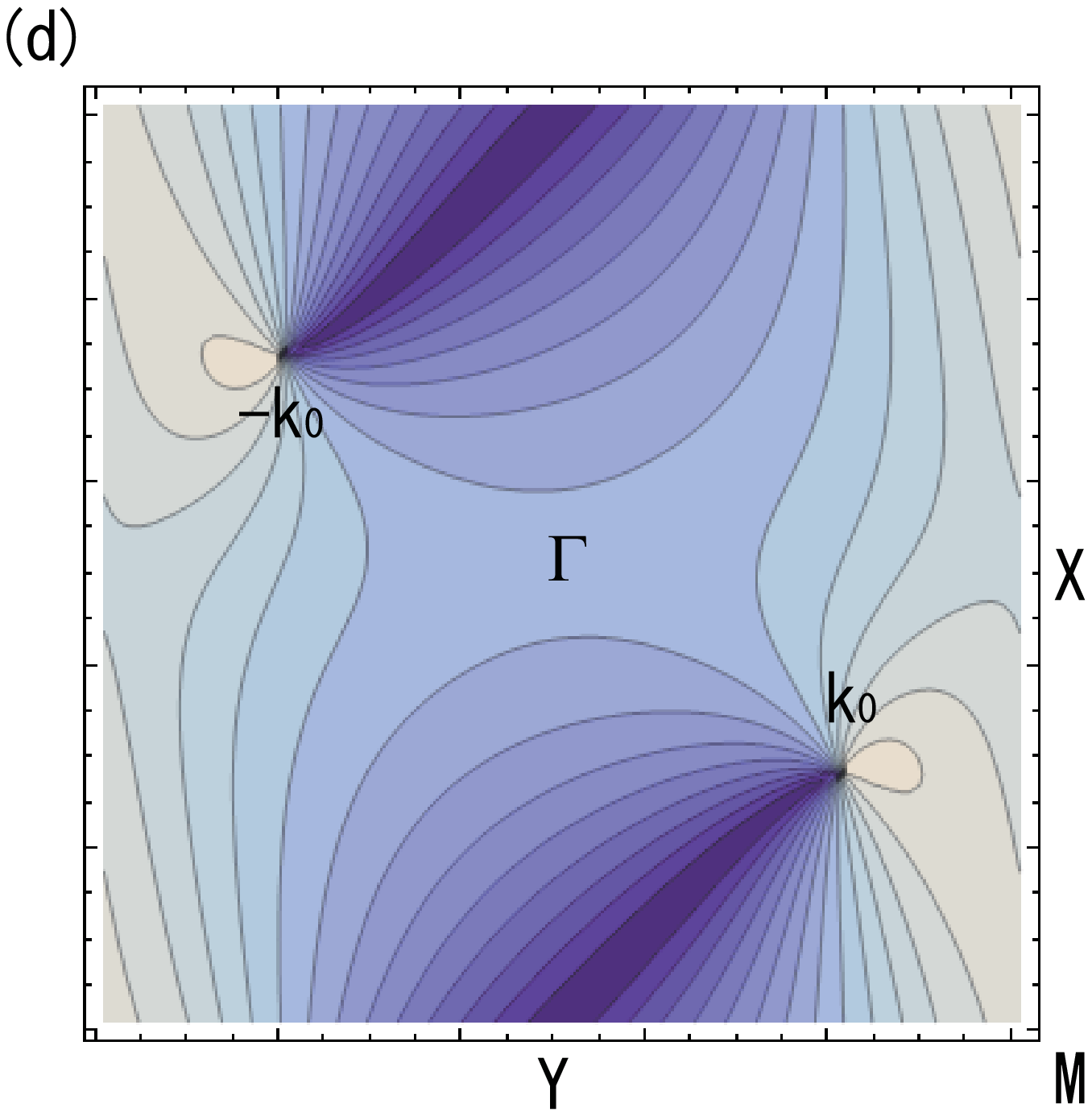}
\includegraphics[height=30mm]{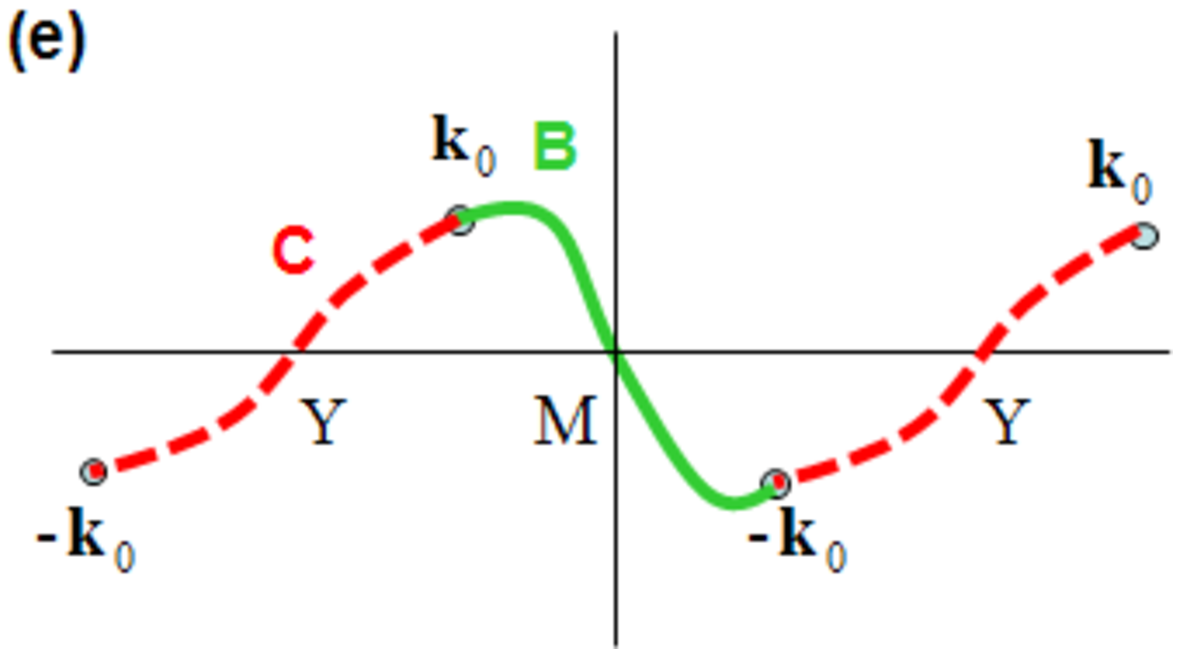}
\caption{\label{fig:zerolines}
Absolute values of the wave functions of the conduction band ($\xi =1$) 
for the components of $B$ site \SU{( 
$|d_{B 1} ({\bf k})  d_{B 1}^\ast ({\bf k})|$(a) and 
 and their coutour plot(b)) and those of $C$ site ((c) and (d)) 
in the Brillouin zone.
The zerolines of the conduction band 
 are bounded by two Dirac points at $\pm {\bf k}_0$
 as shown in the dark region of  (c) and (d), and in 
 schematic figure (e)  for $B$ site (green solid line) and $C$ site (red dashed line).}
(Color Online)
}
\end{figure}

 In Fig 3(a), the conduction band and the valence band are shown 
The chemical potential with $\omega =0$ is situated on the Dirac points at $T=0$,  while it decreases with increasing temperature owing to the asymmetry of the conduction and valence bands.\cite{Kobayashi2008}
 The  saddle points close to the Fermi energy are seen  at the $M$-point 
   and at $Y$-point, respectively.  
 The saddle points give rise to the Van Hove singularities 
   $\omega=0.012$eV in the conduction band, 
and at $\omega=-0.026$eV in the valence band. Such a singularity appears in 
 the components of density of state at  $A$ and $C$ sites
 as  shown in Fig 3(b), where the energy of the Dirac points is $\omega =0$.
For the component of $B$ site, however, these two Van Hove singularities disappear, 
since the zeroline for $B$ site passes through the location of 
 the saddle point.

\begin{figure}
\includegraphics[height=60mm]{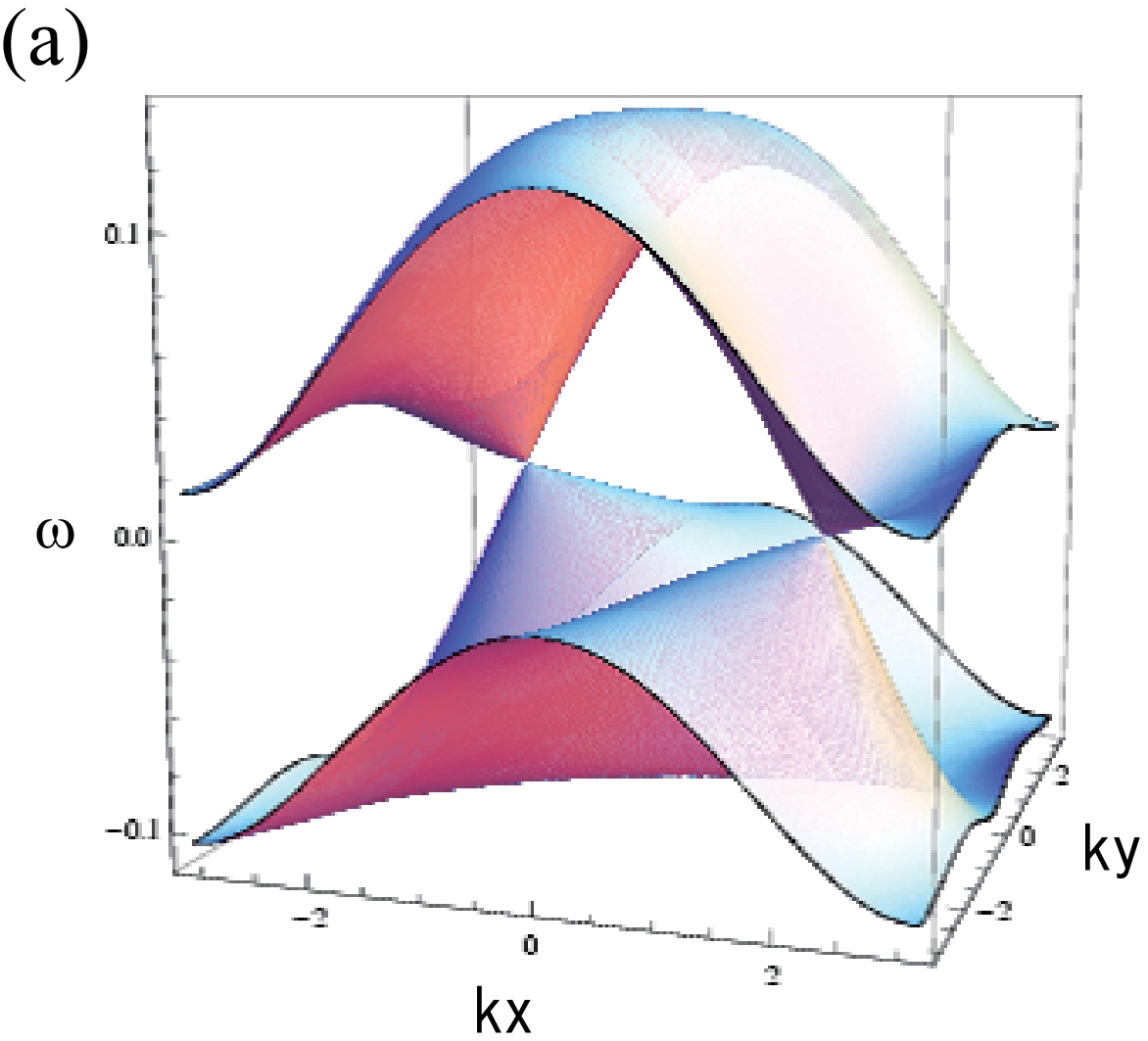}
\includegraphics[height=60mm]{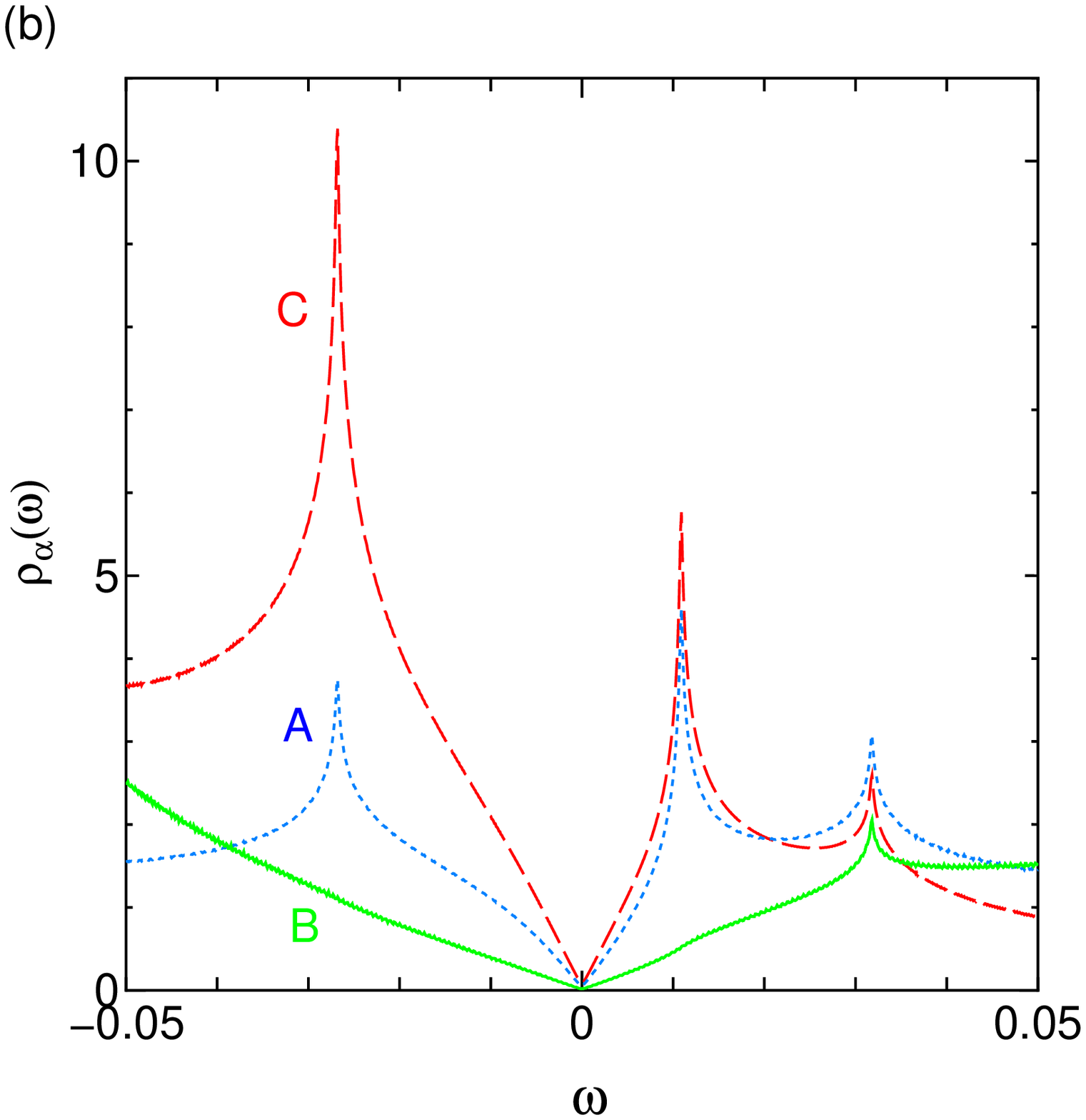}
\caption{\label{fig:band_DOS}
(a) Energy dispersions in the Brillouin zone 
 for the conduction and valence bands ($\gamma =1$ and 2), which has a contact point with the tilted Dirac cones  at $\pm {\bf k}_0$.
The chemical potential is given by with $\omega =0$.
There are the saddle points at the $M$-point in the conduction band 
and at $Y$-point in the valence band near $\omega =0$.
(b) The density of states $\rho_\alpha$ near the energy of the Dirac point ($\omega =0$) at $T=0$ 
for the components of $\alpha = A$ site (blue dotted line), 
$\alpha = B$ site (green solid line), and $\alpha = C$ site (red dashed line).
(Color Online)
}
\end{figure}

\subsection{Knight shift and NMR relaxation rate}

Figure 4(a) shows temperature dependences of the Knight shift $K_\alpha$ 
for $\alpha =A$, $B$, and $C$ sites 
with $U=0.12$ and $K_\alpha^0$ with $U=0$.  
The value of $U=0.12$ is chosen so that the temperature dependences of $K_\alpha$  
 in the present calculation \SU{reproduce}  those of experimental results for $T>50$K.
The Knight shift $K_\alpha^0$ (with $U=0$) for $T<0.015$ has been calculated 
in ref. \onlinecite{Katayama2009}, \SU{which gives slightly different results 
 due to ignoring  the $T$-dependence of the chemical potential.}
In high temperature region ($100 < T < 300$K), 
\SU{With decreasing temperature}, the Knight shift monotonously decreases for $B$-site while there is a maximum in the temperature dependence of the Knight shift for $A$ and $C$ sites.
Such behavior is independent of $U$ originates from 
\SU{the combined effect of the Van Hove singularity and} the zerolines as discussed in Fig 3.
In medium temperature region ($50 < T < 100$K), on the other hand, it is shown that 
the Knight shift is convex downward with decreasing temperature for $B$-site with $U=0.12$, 
while the components for $A$ and $C$ sites exhibit linear $T$-dependences in this region.
\SU{Such a effect of} electron correlation originates from the ferrimagnetic spin fluctuation described later.
Figure 4(b) shows temperature dependences of $(1/T_1 T)_\alpha$  
for $\alpha =A$, $B$, and $C$ sites 
with $U=0.12$ and $(1/T_1 T)_\alpha^0$ with $U=0$.  
It is shown that $(1/T_1 T)_\alpha$ exhibit linear $T^2$-dependences with $T < 100K$ approximately, 
and all components are weakly enhanced by $U$.
Figure 4(c) shows temperature dependences of the Korringa ratio $(1/T_1 T K^2)_\alpha$  
for $\alpha =A$, $B$, and $C$ sites 
with $U=0.12$ and $(1/T_1 T K^2)_\alpha^0$ with $U=0$.  
It is clearly shown that the Korringa ratio for $B$ site is strongly enhanced by $U$ 
and increases with decreasing $T$, owing to the anomalous $T$-dependence of $K_{\rm B}$.   
The electron correlation effect results in 
 \SU{the enhancement of the following inequality,}  
\begin{equation}
(1/T_1 T K^2)_{\rm B} > (1/T_1 T K^2)_{\rm A} > (1/T_1 T K^2)_{\rm C} .
\end{equation}

\begin{figure}
\includegraphics[height=60mm]{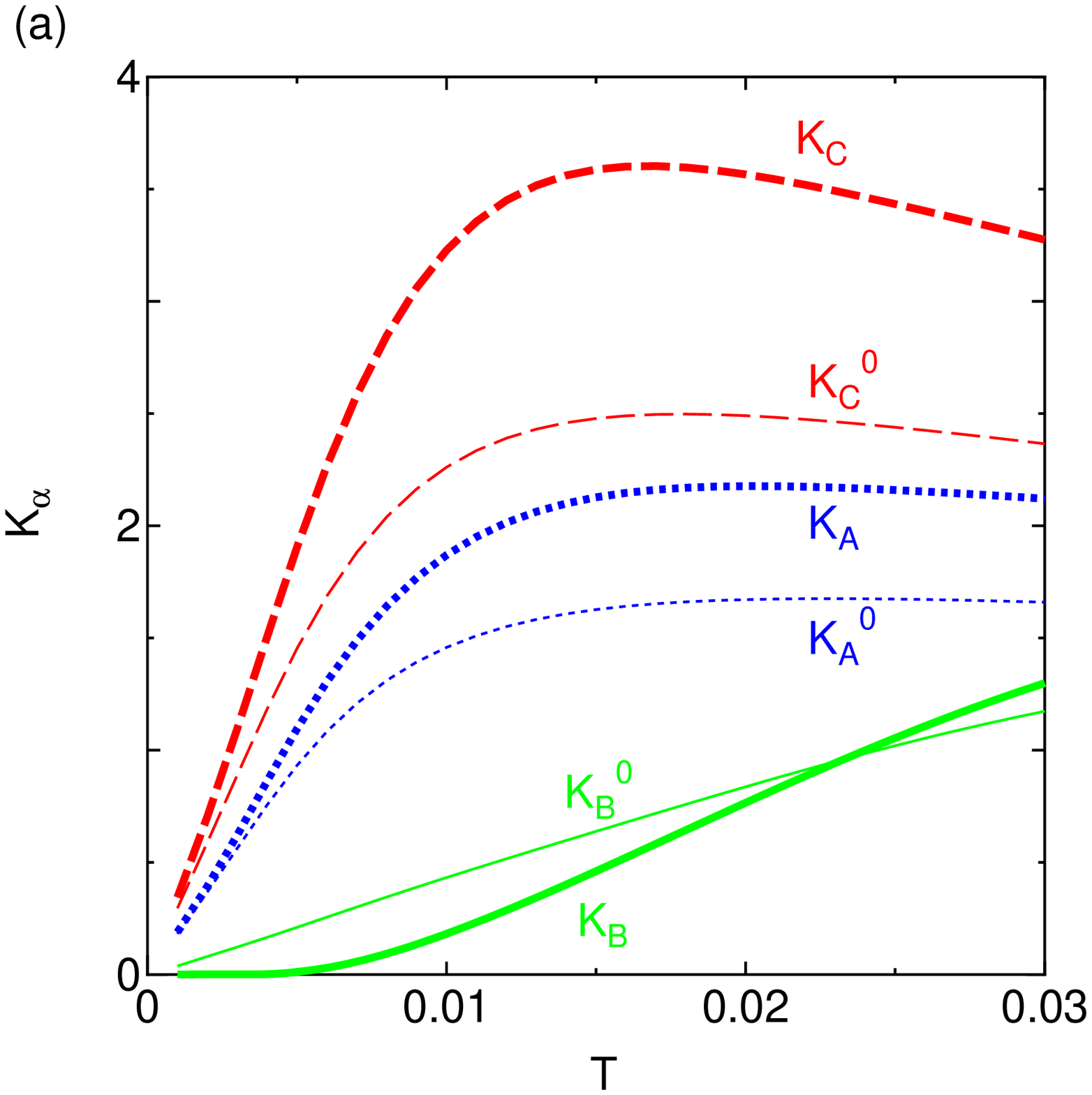}
\includegraphics[height=60mm]{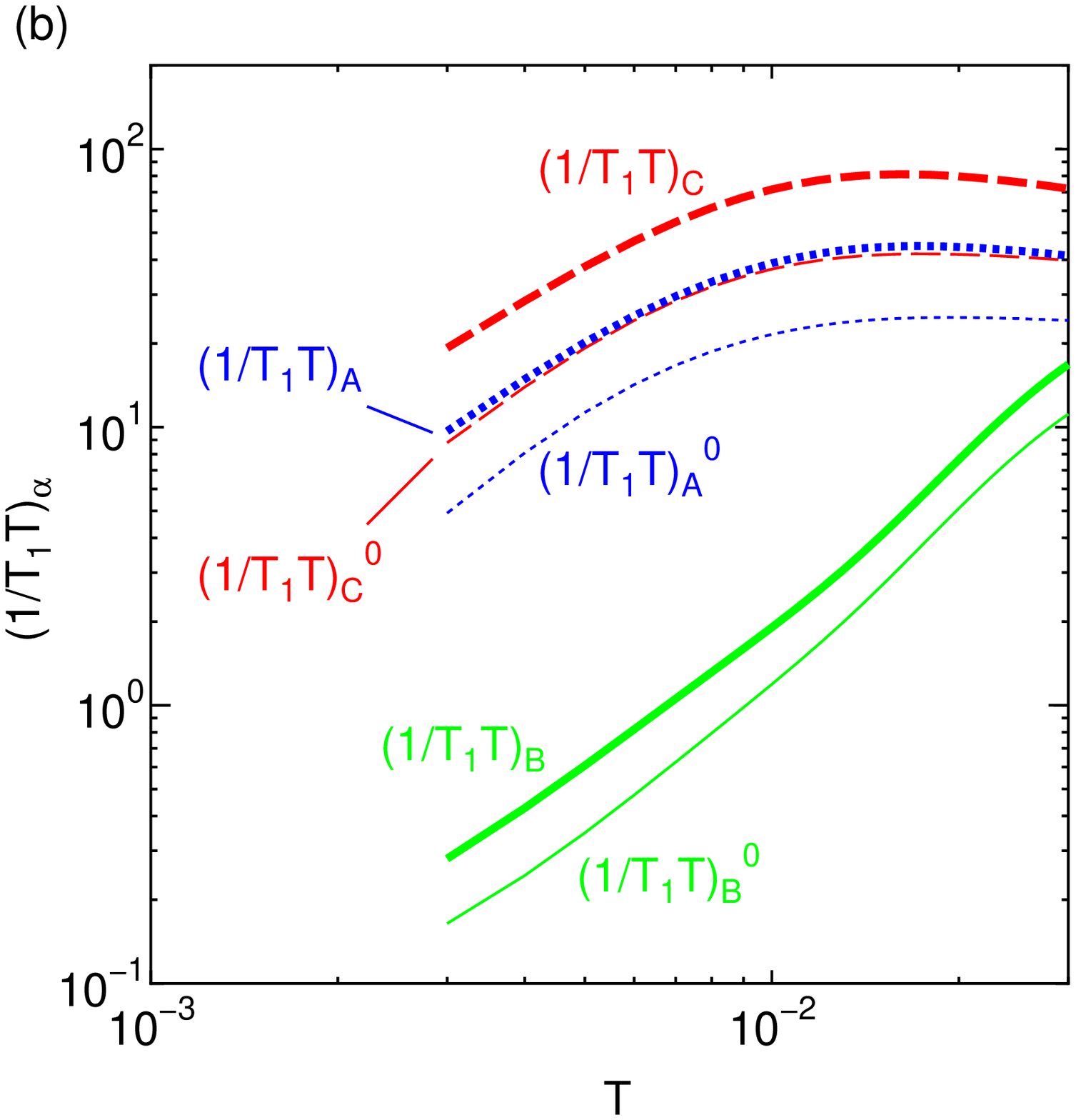}
\includegraphics[height=60mm]{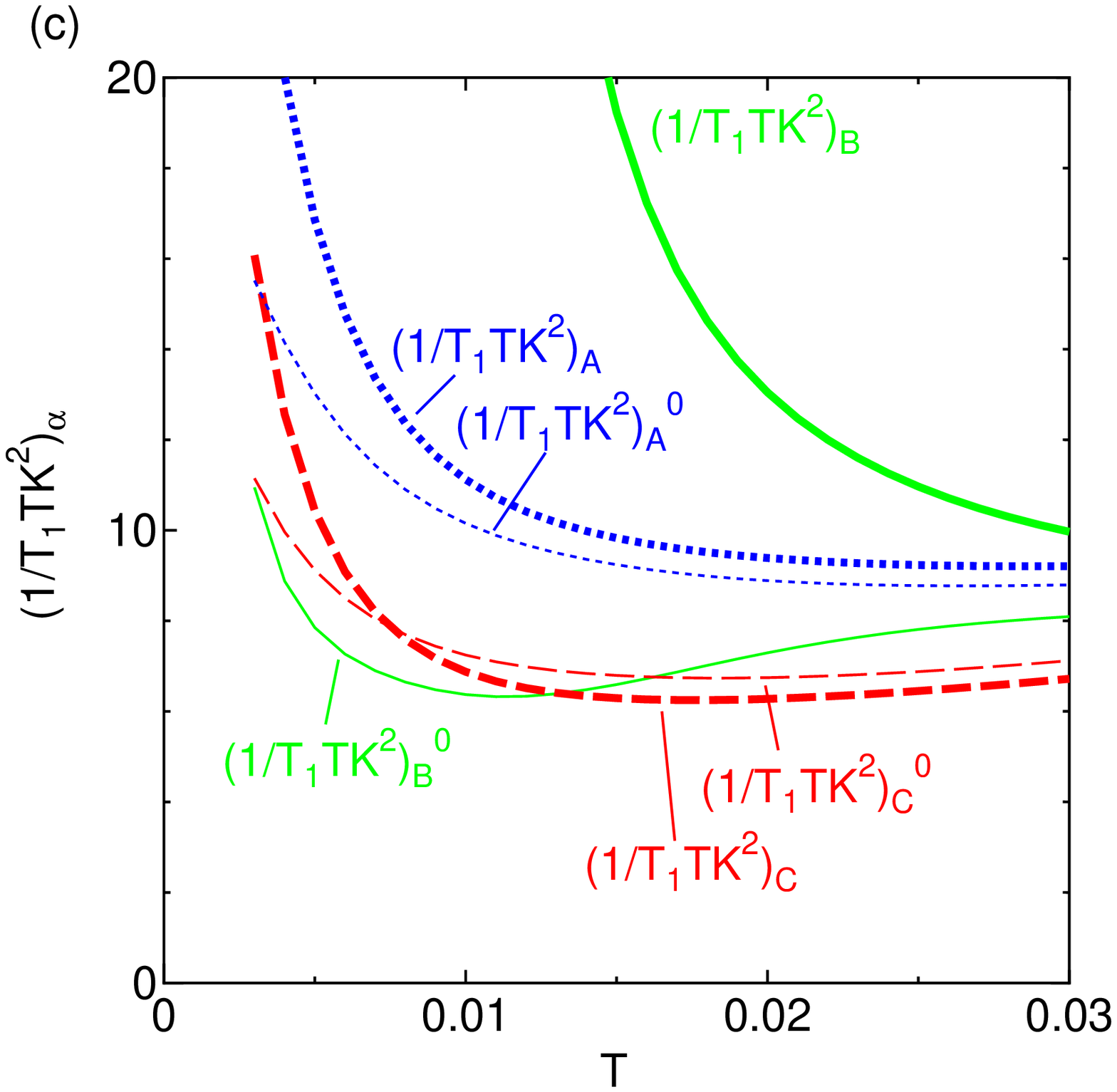} 
\caption{\label{fig:band_DOS}
(a) Temperature dependences of the Knight shift $K_\alpha$  with $U=0.12$ (thick line) and $K_\alpha^0$ (thin line)
for $\alpha =A$ (red dashed line), $B$ (green solid line), 
and $C$ sites (blue dotted  line), respectively.
(b) Temperature dependences of $(1/T_1 T)_\alpha$ with $U=0.12$ (thick line) 
 and $(1/T_1 T)_\alpha^0$ (thin line)
 for $\alpha =A$ (red dashed line), $B$ (green solid line), 
and $C$ sites (blue dotted  line), respectively.  
(c) Temperature dependences of Korringa ratio $(1/T_1 T K^2)_\alpha$ with $U=0.12$ (thick line) and $(1/T_1 T K^2)_\alpha^0$ with $U=0$ (thin line)
 for $\alpha =A$ (red dashed line), $B$ (green solid line), 
and $C$ sites (blue dotted line), respectively. 
(Color Online)
}
\end{figure}

\subsection{Ferrimagnetic fluctuation in spin susceptibility}

\SU{ We examine static spin susceptibility at low energy  by choosing
   $T=0.02$ and $\omega =0$. 
In Fig 5(a), the momentum dependences of the diagonal components of the spin susceptibilities $\chi_{\alpha \alpha}^s$ 
  and the bare susceptibilities $\chi_{\alpha \alpha}^0$  ($U=0$)
 for $\alpha =A$, $B$, and $C$  are shown}  
for $\alpha =A$, $B$, and $C$ with $U=0.12$.
The diagonal components are positive in the Brillouin zone and are enhanced by $U$.
The momentum dependences of the off-diagonal components of the spin susceptibilities 
$\chi_{\alpha \beta}^s$  
for $(\alpha ,\beta ) =(A, B)$, $(B, C)$, and $(A, C)$ with $U=0.12$, 
and the bare susceptibilities $\chi_{\alpha \beta}^0$ for $(\alpha ,\beta ) =(A, B)$,  
$(B, C)$, and $(A, C)$ with $U=0$ are shown in Fig 5(b). 
It is \SU{found}  that $\chi_{\rm BC}^s$ and $\chi_{\rm AB}^s$ are negative in the Brillouin zone, 
while $\chi_{\rm AC}^s$ has both positive and negative value.
Those absolute values are enhanced by $U$.
Especially, $\chi_{\rm BC}^s$ plays the most important role for the anomalous $T$-dependences
 of $K_{\rm B}$, since $\vert \chi_{\rm BC}^s \vert $ at ${\bf q} ={\bf 0}$ is strongly enhanced by $U$, 
owing to product of negative $\chi_{\rm BC}^0$ with positive diagonal terms 
in the RPA process such as

\begin{eqnarray}
[ \hat{\chi}^s ]_{\rm BC} = [ (1-\hat{\chi}^0 \hat{U})^{-1} \hat{\chi}^0 ]_{\rm BC} \nonumber \\
=\chi_{\rm BC}^0 +U \chi_{\rm BB}^0 \chi_{\rm BC}^0 +U \chi_{\rm BC}^0 \chi_{\rm CC}^0 +\cdots . 
\end{eqnarray}

Negative off-diagonal susceptibilities, $\chi_{\rm BC}^s$ and $\chi_{\rm AB}^s$, 
and positive off-diagonal susceptibility, $\chi_{\rm AC}^s$, at ${\bf q} ={\bf 0}$ 
indicate the ferrimagnetic spin fluctuation where the spin on B site tends to be opposite 
to that of the other sites as shown in Fig 5(c).
The Knight shift is a sum of the diagonal and off-diagonal spin susceptibilities, 
where the former are dominant at high temperatures since the spin fluctuation exhibits 
local character.
The latter become important with decreasing temperature, resulting in the anomalous $T$-dependence 
of $K_{\rm B}$ which is convex downward at low temperatures.

\begin{figure}
\includegraphics[height=60mm]{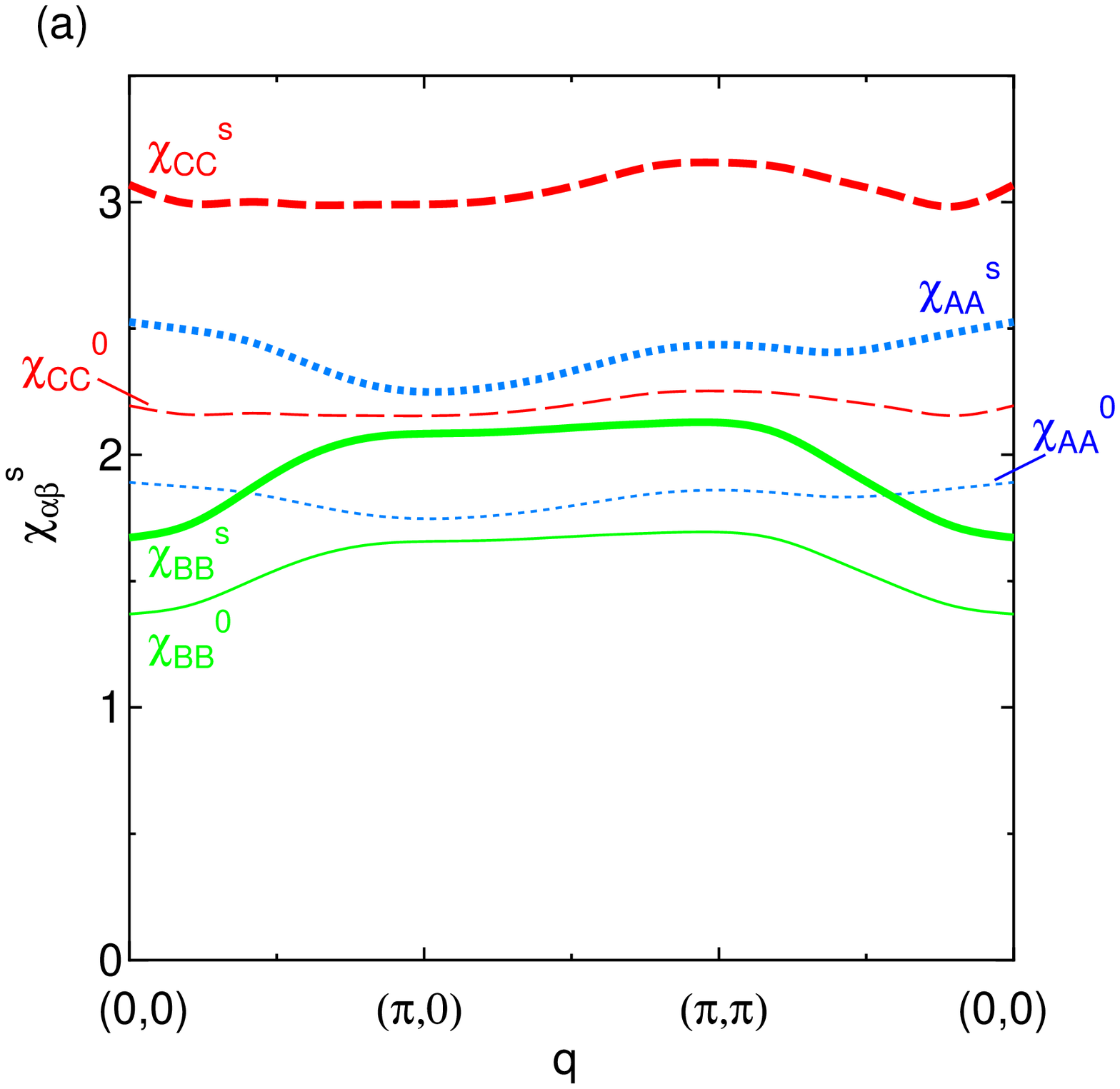}
\includegraphics[height=60mm]{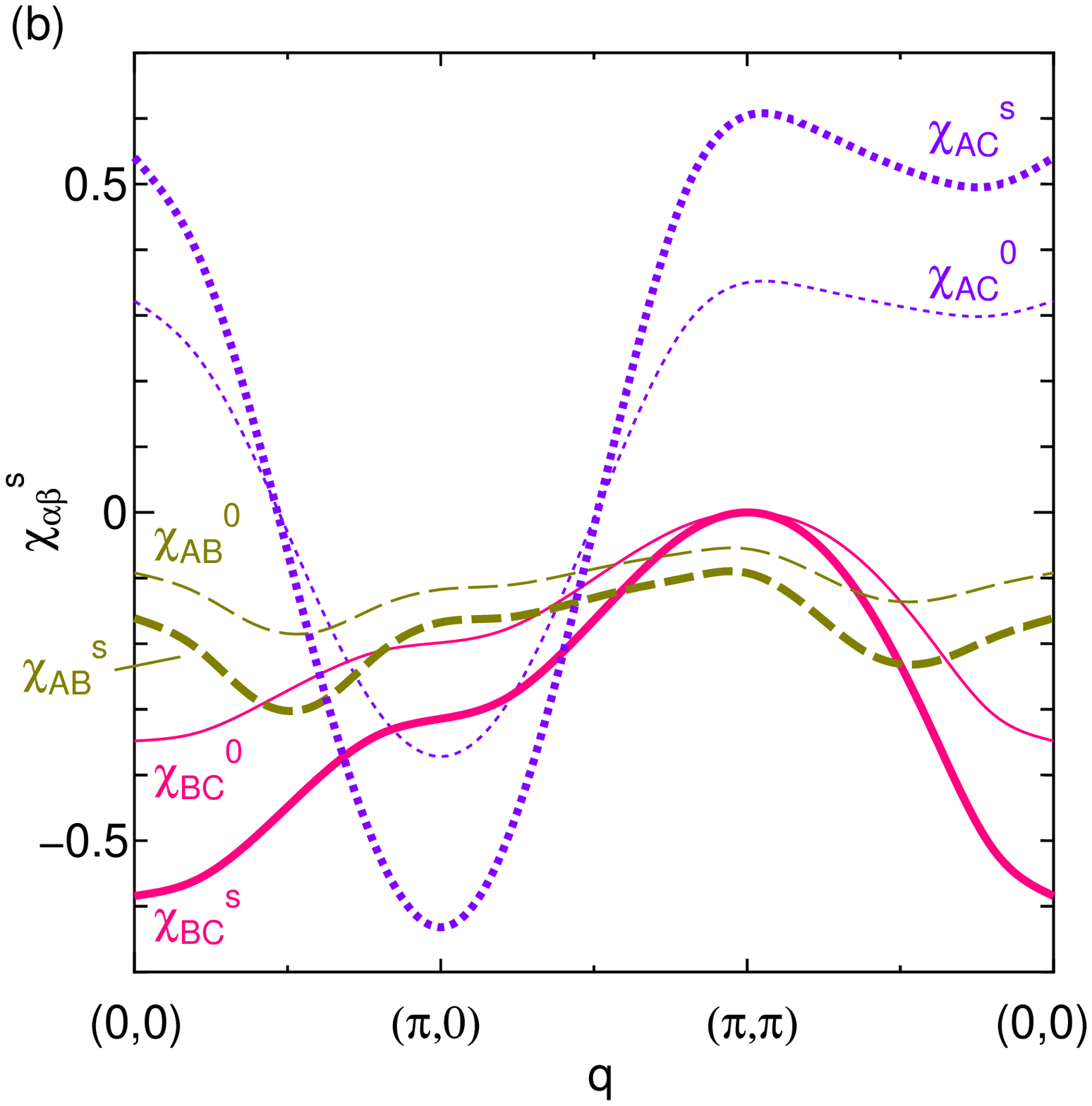}
\includegraphics[height=40mm]{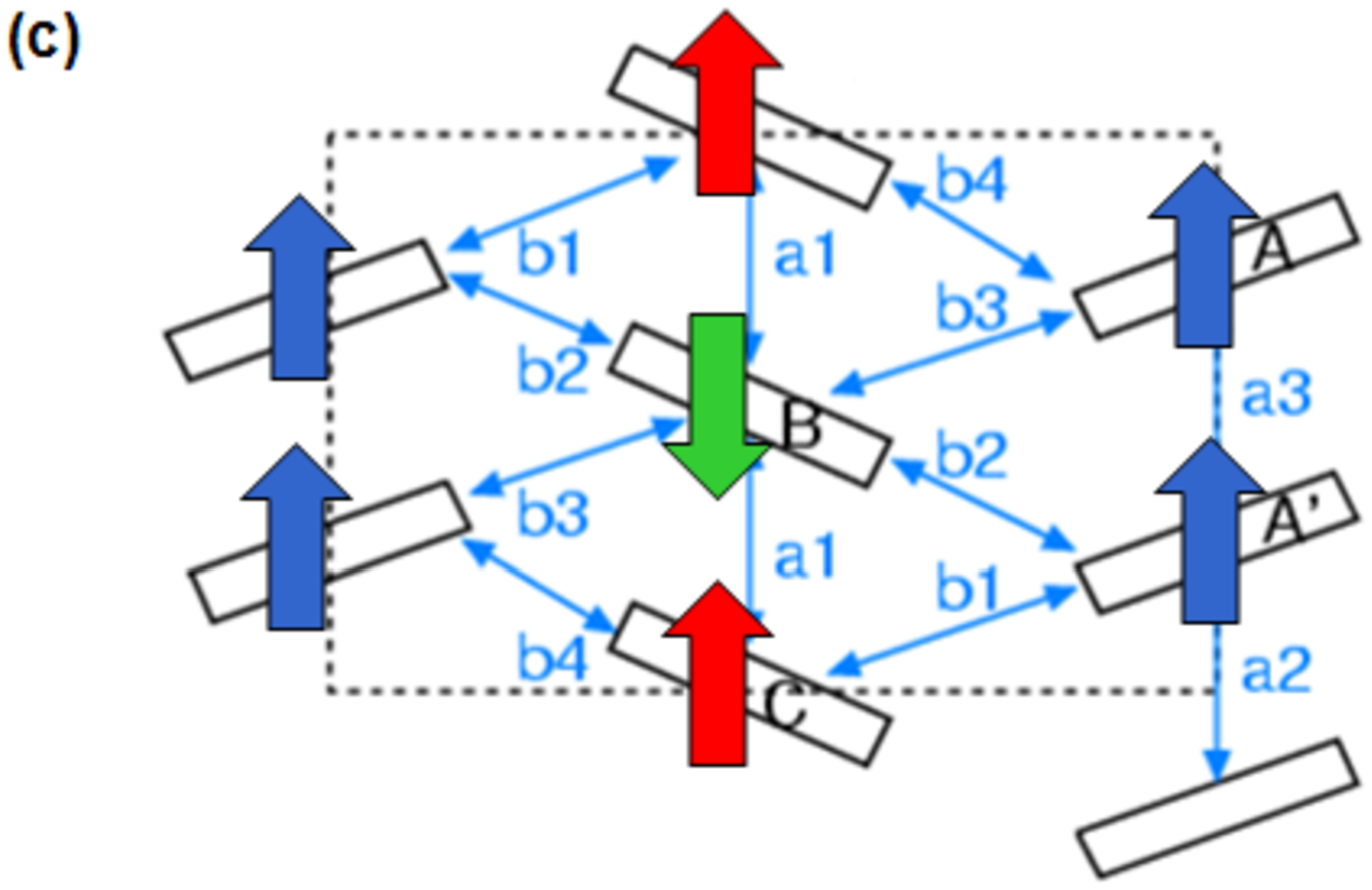}
\caption{\label{fig:band_DOS}
(a) Momentum dependences of the diagonal components of the spin susceptibilities $\chi_{\alpha \alpha}^s$ with $U=0.12$ (thick line) and  
 the bare susceptibilities $\chi_{\alpha \beta}^0$ (thin line) 
  for $\alpha =A$ (red dashed line), $B$ (green solid  line), 
and $C$ sites (blue dotted line), 
where $T=0.02$ and $\omega =0$. 
(b) Momentum dependences of the off-diagonal components of the spin susceptibilities 
$\chi_{\alpha \beta}^s$ $U=0.12$ (thick line) and 
  the bare susceptibilities $\chi_{\alpha \beta}^0$ (thin line)
 for $(\alpha ,\beta ) =(A, B)$ (gold dashed line), $(B, C)$ 
(wine red solid line), and $(A, C)$ sites (purple dotted line),
where $T=0.02$ and $\omega =0$.
(c) Schematic figure for pattern of the ferrimagnetic spin fluctuation.
(Color Online)
}
\end{figure}

The negative value of $\chi_{\rm BC}^0$ at ${\bf q}={\bf 0}$ originats from the form factor 
for the interband fluctuation, $F_{\rm BC}^{(2)} ({\bf k} ,{\bf 0})$,  
since the form factor for intraband fluctuation, $F_{\rm BC}^{(1)} ({\bf k} ,{\bf 0})$, 
is positive for any ${\bf k}$.
The momentum ${\bf k}$ dependence of $F_{\rm BC}^{(2)} ({\bf k} ,{\bf 0})$ in the Brillouin zone 
is shown in Figs 6(a) and 6(b).
Although $F_{\rm BC}^{(2)} ({\bf k} ,{\bf 0})$ can take positive or negative values, 
the numerical result show $F_{\rm BC}^{(2)} ({\bf k} ,{\bf 0})$ is negative or zero
for any ${\bf k}$.
It is not a self-evident result owing to the phase structure of the wave function as discussed later.
The white curved lines in Fig. 6(b) correspond to the zeroline for B and C sites, 
where $F_{\rm BC}^{(2)} ({\bf k} ,{\bf 0})=0$.

\begin{figure}
\includegraphics[height=60mm]{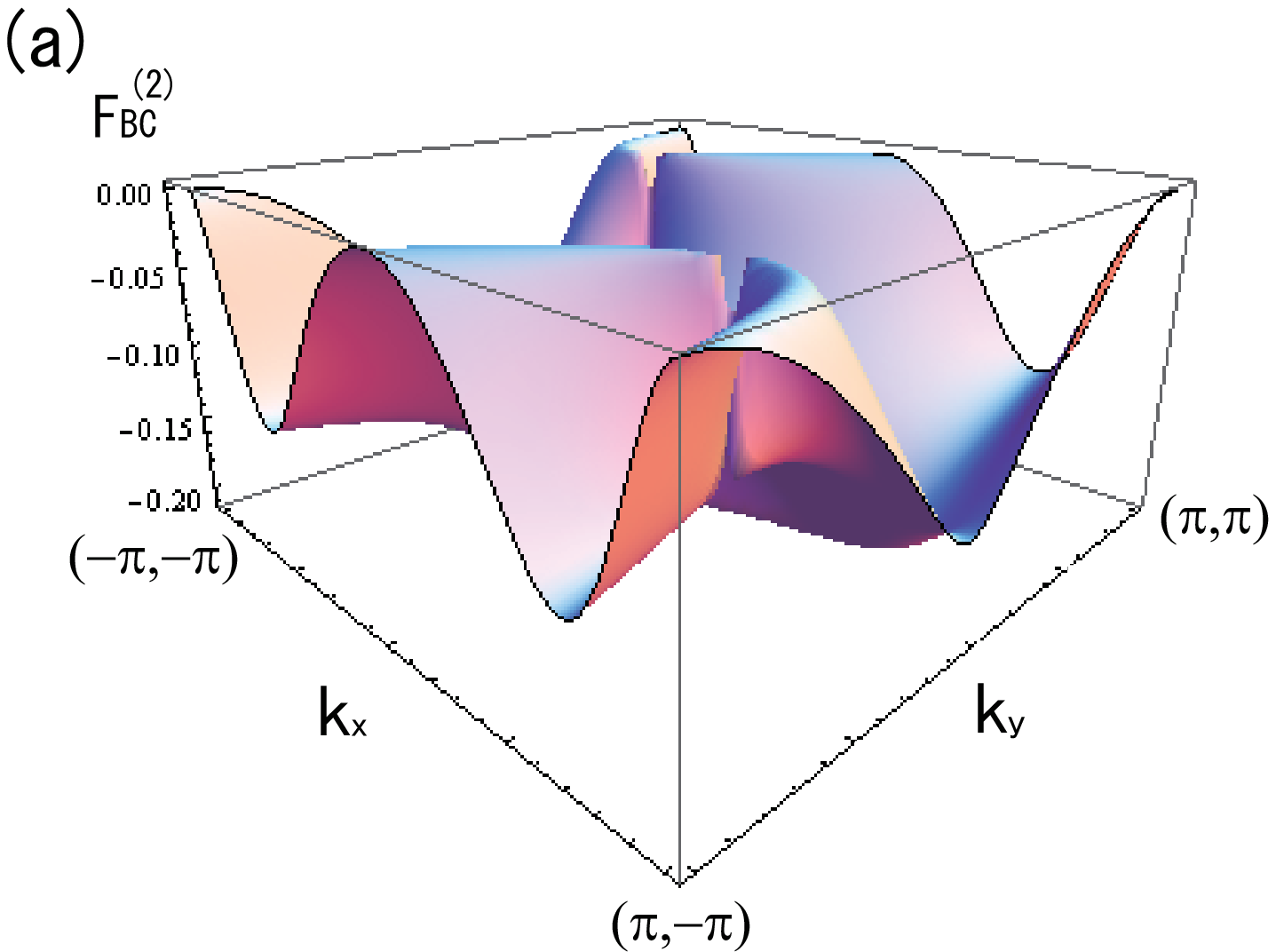}
\includegraphics[height=60mm]{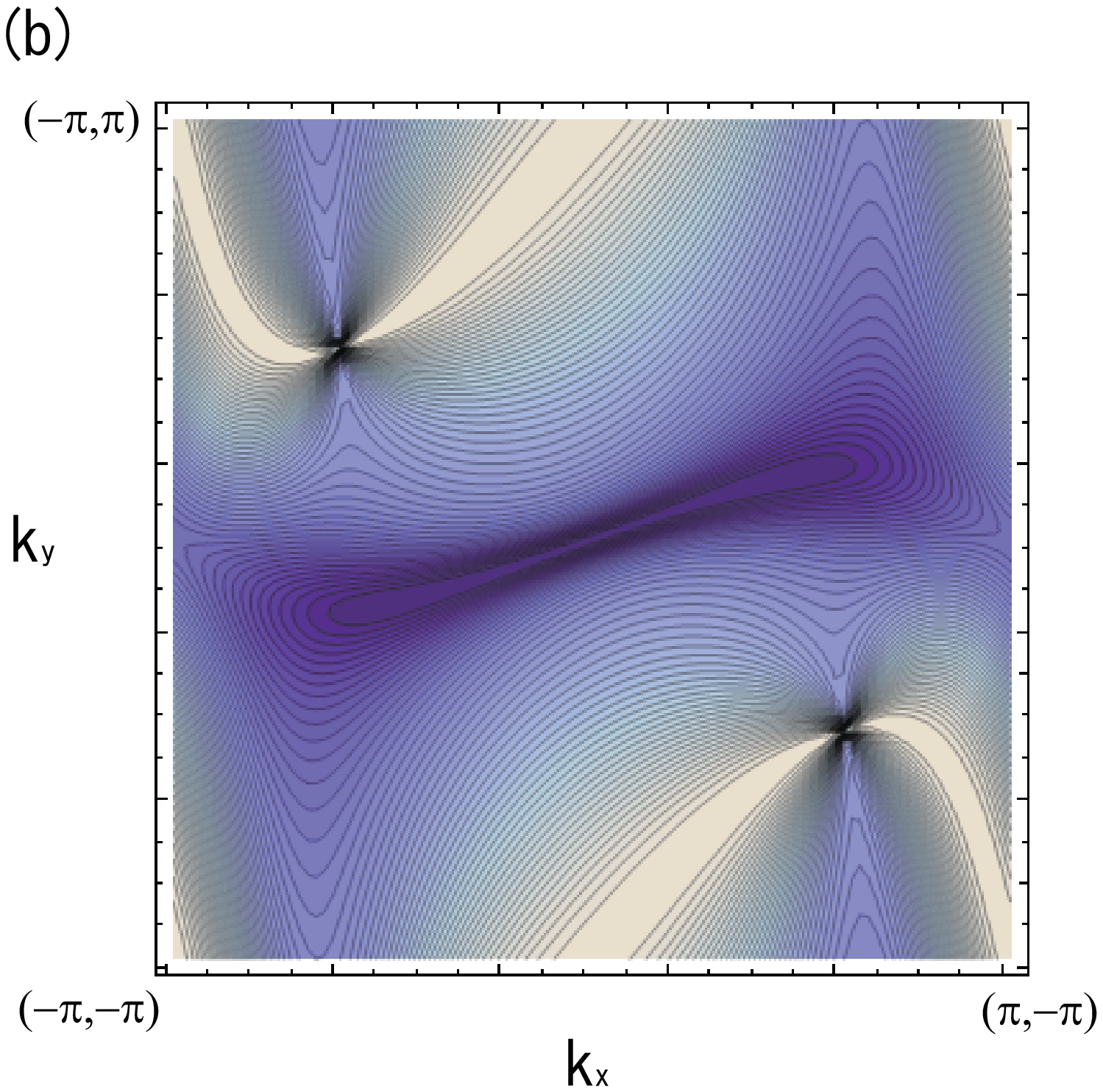}
\caption{\label{fig:band_DOS}
(a) Momentum ${\bf k}$ dependence of the form factor
\SU{ $F_{\rm BC}^{(2)} ({\bf k} ,{\bf 0})$}
 for the interband fluctuation
in the Brillouin zone.
The contour plot \SU{corresponding to (a)} is shown in (b)
(Color Online)
}
\end{figure}


In order to show the reason of $F_{\rm BC}^{(2)} ({\bf k} ,{\bf 0}) \le 0$, 
we investigate the phase structure of the wave function.
Properties of wave functions in the vicinity of the Dirac points have been investigated \SU{ by analying in terms of} 
 the Luttinger-Kohn representation\cite{Katayama2009}.
The analysis of wave functions in the present paper, on the other hand, 
is based on the Bloch representation.
Figure 7 shows numerical results of wave functions 
for B and C sites and for conduction and valence bands,  
$d_{{\rm B}1} ({\rm k_0}^\prime)^\ast$, 
$d_{{\rm B}2} ({\rm k_0}^\prime)$,
$d_{{\rm C}1} ({\rm k_0}^\prime)$ and 
$d_{{\rm C}2} ({\rm k_0}^\prime)^\ast$, 
on the Gauss plane, \SU{here $\vert {\rm k_0}^\prime -{\rm k_0} \vert =0.05 \pi$,
i.e., the momentum ${\rm k_0}^\prime$ circles the Dirac point ${\bf k}_0$.}
\SU{In order to avoid arbitrary phase factor of the wave functions 
for each band $\gamma$ and each momentum ${\rm k_0}^\prime$, 
we choose $d_{{\rm A} \gamma} ({\rm k_0}^\prime)$ is real and positive.
Such a representation for wave functions was used  in ref. \onlinecite{Katayama} to examine the rotation of the base, $d_{\alpha,1}$ and  $d_{\alpha,2}$,  around  ${\bf k}_0$.}
\SU{The inset shows $\theta$-dependences of absolute values of the wave functions for the conduction band. 
The similar result was shown in \onlinecite{Katayama2009} although there is slight difference in the choice of the horizontal axis.}
Based on this numerical results, the wave functions
in the vicinity of the Dirac points, 
 are approximately given by 
\begin{eqnarray}
d_{{\rm B}1} (\theta ) &=& e^{{\rm i} \theta_1} \vert d_{{\rm B}1} \vert 
[ e^{{\rm i} (\theta +\theta_{{\rm B}1})} - e^{{\rm i} (\theta_{\rm BZ} +\theta_{{\rm B}1})} ], 
\nonumber \\
d_{{\rm C}1} (\theta ) &=& e^{{\rm i} \theta_1} \vert d_{{\rm C}1} \vert 
[ e^{{\rm i} (\theta +\theta_{{\rm C}1})} - e^{{\rm i} (\theta_{\rm CZ} +\theta_{{\rm C}1})} ], 
\nonumber \\
d_{{\rm B}2} (\theta ) &=& e^{{\rm i} \theta_2} \vert d_{{\rm B}2} \vert 
[ e^{{\rm i} (\theta +\theta_{{\rm B}2})} - e^{{\rm i} (\theta_{\rm CZ} +\theta_{{\rm B}2})} ], 
\nonumber \\
d_{{\rm C}2} (\theta ) &=& e^{{\rm i} \theta_1} \vert d_{{\rm C}2} \vert 
[ e^{{\rm i} (\theta +\theta_{{\rm C}2})} - e^{{\rm i} (\theta_{\rm BZ} +\theta_{{\rm C}2})} ], 
\end{eqnarray}
where $\theta$ is the angle of the vector ${\rm k_0}^\prime -{\rm k_0}$ measured from $k_x$-axis.
The values of wave functions at $\theta =0$ are determined by 
$\theta_{{\rm B}1}$, $\theta_{{\rm C}1}$, $\theta_{{\rm B}2}$ and $\theta_{{\rm C}2}$.
The angle of the zeroline for B (C) site in the conduction band is 
$\theta_{\rm BZ}$ ($\theta_{\rm CZ}$), 
where $\vert d_{{\rm B}1} (\theta ) \vert =0$ ($\vert d_{{\rm C}1} (\theta ) \vert$ =0) 
as shown in the inset of Fig. 7, 
while the zeroline for B (C) site in the valence band corresponds to 
$\theta_{\rm CZ}$ ($\theta_{\rm BZ}$).
The arbitrary phases of the wave functions for each band and each momentum, 
$\theta_1 (\theta )$ and $\theta_2 (\theta )$, disappear in the form factor.
Using above wave functions with the parameters given by the numerical calculation, 
we obtain $F_{\rm BC}^{(2)} ({\bf k}^0 ,{\bf 0}) \le 0$ at arbitrary $\theta$.

If $\theta_{\rm CZ} =\theta_{\rm BZ} +\pi$ (close to the present numerical result), 
\begin{equation}
F_{\rm BC}^{(2)} ({\bf k}^0 ,{\bf 0}) =-8 \cos \theta_0 \sin^2 (\theta -\theta_{\rm BZ} )
\end{equation}
with $\theta_0 =-\theta_{{\rm B}1} +\theta_{{\rm C}1} -\theta_{{\rm C}2} +\theta_{{\rm B}2}$.
We obtain $F_{\rm BC}^{(2)} ({\bf k}^0 ,{\bf 0}) \le 0$ in the wide region of parameter, 
$-\pi /2 \le \theta_0 \le \pi /2$.
Thus the condition of $F_{\rm BC}^{(2)} ({\bf k}^0 ,{\bf 0}) \le 0$ is robust 
when the zerolines of B and C sites extend in opposite directions each other.
When $\theta_{\rm CZ} =\theta_{\rm BZ}$, on the other hand, 
\begin{equation}
F_{\rm BC}^{(2)} ({\bf k}^0 ,{\bf 0}) =8 \cos \theta_0 (1-\cos (\theta -\theta_{\rm BZ} ))^2 .
\end{equation}
We obtain $F_{\rm BC}^{(2)} ({\bf k}^0 ,{\bf 0}) \le 0$ with $\theta_0$ in the present numerical result.
Thus the angle between the zerolines of B and C sites is an important factor 
for the ferrimagnetic fluctuation in the Dirac electrons of $\alpha$-(BEDT-TTF)$_2$I$_3$.

\begin{figure}
\includegraphics[height=60mm]{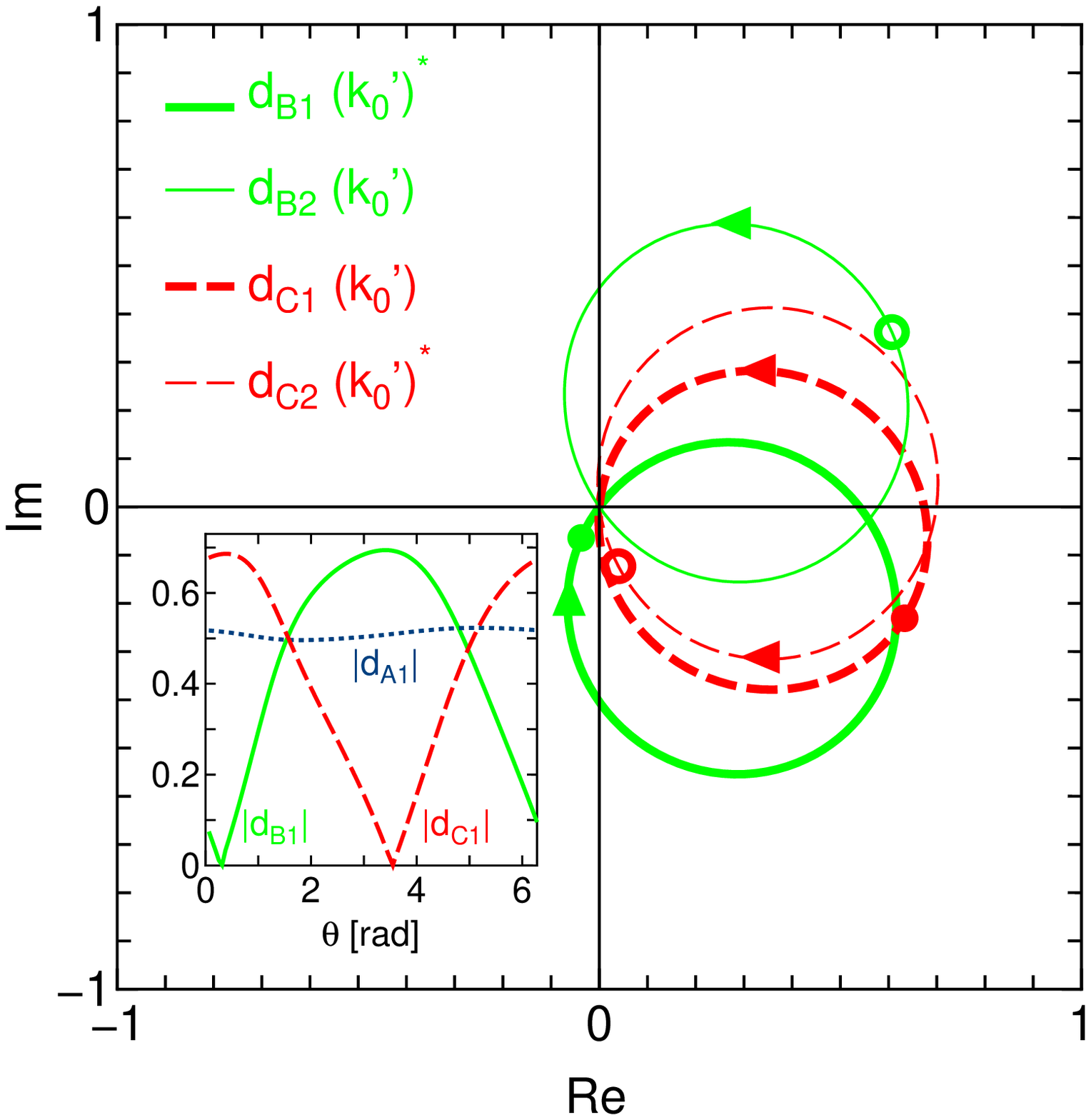}
\caption{\label{fig:band_DOS}
Numerical results of 
Wave functions normalized by  $d_{{\rm A}\gamma}/|d_{{\rm A}\gamma1}|$
 ($\gamma=$1,2) on the Gauss plane, 
for $d_{{\rm B}1} ({\rm k_0}^\prime)^\ast$ (the green thick solid circle), 
$d_{{\rm B}2} ({\rm k_0}^\prime)$ (the green thin solid circle),
$d_{{\rm C}1} ({\rm k_0}^\prime)$ (the red thick dashed circle) and 
$d_{{\rm C}2} ({\rm k_0}^\prime)^\ast$ (the red thin dashed circle), 
where $\vert {\rm k_0}^\prime -{\rm k_0} \vert =0.05 \pi$ and the angle 
of ${\rm k_0}^\prime -{\rm k_0}$ vector from $k_x$-axis, $\theta$, moves from zero to $2 \pi$.
The directions where the values of wave functions move with increasing $\theta$ are 
represented by the arrows on the circles, 
and the values of wave functions at $\theta =0$ are indicated by the filled or open small circles 
on the large circles.
The inset shows $\theta$-dependences of absolute values of the wave functions for the conduction band, 
$\vert d_{{\rm A}1} ({\rm k_0}^\prime) \vert$ (the blue dotted line), 
$\vert d_{{\rm B}1} ({\rm k_0}^\prime) \vert$ (the green solid line) and 
$\vert d_{{\rm C}1} ({\rm k_0}^\prime) \vert$ (the red dashed line.
(Color Online)
}
\end{figure}

\section{Summary and discussion}

\SU{In summary, we examined the wave function 
  and  the spin fluctuation in $\alpha$-(BEDT-TTF)$_2$I$_3$,
using a tight-binding model and the on-site Coulomb interaction 
 treated within the random phase approximation. 
The effect of electron correlation   on
 $K_\alpha$ and $(1/T_1 T)_\alpha$ 
with ${\rm A}$, ${\rm A}^\prime$, ${\rm B}$ and ${\rm C}$ sites 
for $T > 50$K was investigated 
paying attention to the inequivalence of these sites 
(${\rm A} ={\rm A}^\prime \ne {\rm B} \ne {\rm C}$).}

We found that zerolines, where the wave function is zero for the components of $B$ or $C$ site. They give the vanishing of two Van Hove singularities near the energy of the Dirac points 
only for $B$-site component of DOS.
Existence (absence) of the Van Hove singularities plays essential role 
for the  $T$-dependences of $K_\alpha$.
\SU{For}  the high temperature region of $100 < T < 300$K, 
with decreasing $T$, $K_{\rm B}$ decreases monotonously  
while $K_{\rm A}$ and $K_{\rm C}$ exhibit a maximum.
In the region for $50 < T < 100$K,  
$K_{\rm B}$ is convex downward, and $(1/T_1 T K^2)_{\rm B}$ increases
 leading to an inequality of the Korringa ratio, 
$(1/T_1 T K^2)_{\rm B} > (1/T_1 T K^2)_{\rm A} > (1/T_1 T K^2)_{\rm C}$, was obtained.
These results are consistent with those of experiment for $T > 50$K 
\cite{TakahashiNMR,KanodaNMR1,KanodaNMR2} .

\SU{It is found  that the anomalous $T$-dependence of $K_{\rm B}$  
 is ascribed to the ferrimagnetic spin fluctuation which is} enhanced by 
the on-site Coulomb interaction. Such a fluctuation describes  a spin on ${\rm B}$ site being  opposite to 
the other spins on ${\rm A}$, ${\rm A}^\prime$ and ${\rm C}$ sites.
The ferrimagnetic spin fluctuation originates from  
the interband fluctuation mainly between ${\rm B}$ and ${\rm C}$ sites.
The interband fluctuation relates to the zerolines 
bounded by two Dirac points in the Brillouin zone.
Such zeroline does not exist in the wave function of graphene, 
since carbon atoms in two sublattice are equivalent owing to the inversion symmetry, 
which corresponds to ${\rm A}$ and ${\rm A}^\prime$ sites in $\alpha$-(BEDT-TTF)$_2$I$_3$.
Thus the present results reveal that the inequivalence of BEDT-TTF sites 
play important roles for observables in NMR 
as an inner degree of freedom of Dirac electron in molecular conductor 
with the short range Coulomb interaction.

There remain problems to be clarified.
In the low temperature region for $T < 50$,  
$K_\alpha$ for all components are convex downward with decreasing $T$ 
\cite{TakahashiNMR,KanodaNMR1,KanodaNMR2} , 
and $(1/T_1 T)_\alpha$ exhibits complex $T$-dependence for very low temperatures 
\cite{KanodaNMR2,ShimizuNMR} in the presence of magnetic field perpendicular to 
the conducting plane.
Those behavior can not be explained within the random phase approximation 
on the on-site Coulomb interaction.
It indicates an importance of  a higher order correction such as self-energy correction 
with the long range Coulomb interaction, \SU{since} 
 at low temperatures,  low energy phenomena in the vicinity of Dirac point is dominant and then scale of length is much longer than lattice constant.
\SU{Finally, we note that} the valley splitting owing to the pseudo-spin XY ferromagnetism in $N=0$ Landau states 
\cite{Kobayashi2009QHE}
may also play significant roles at very low temperatures 
in the presence of magnetic field perpendicular to the conducting plane.

\acknowledgements
The authors are thankful to K. Ishikawa, M. Hirata, K. Miyagawa and K. Kanoda for fruitful discussions.
\SU{Y.S. is indebted to the Daiko Foundation for financial aid
 in the present work.}
This work was financially supported in part
 by Grant-in-Aid for Special Coordination Funds for Promoting
Science and Technology (SCF), Scientific Research on Innovative
Areas 20110002, and was also financially supported by a Grant-in-Aid for Special Coordination Funds
for Promoting Science and Technology (SCF) from the Ministry of Education, Culture, Sports, Science
and Technology in Japan,
and Scientific Research 19740205, 22540366, 23540403 and 24244053 from the Ministry of
Education, Culture, Sports, Science and Technology in Japan.

\end{document}